\documentstyle[12pt,psfig]{article}
\normalsize
\begin{document}
\noindent{
{\Large\bf LANL Report LA-UR-99-360 (1999)\\
}
}
Submitted to {\em Nucl. Instr. Meth. A} and to be presented 
at the {\em American Physical Society Centennial Meeting,
Atlanta, Georgia, March 20-26, 1999}
\vspace*{0.5cm}

\begin{center}
{\Large\bf
Computer  Study of Isotope Production\\ in High Power Accelerators\\
}
\vspace*{0.5cm}
{ K.~A.~Van Riper$^{a}$, S.~G.~Mashnik$^{b}$, and W.~B.~Wilson$^{b}$}\\
{\em $^{a}$ White Rock Science, PO Box 4729, Los Alamos, NM 87544}\\
{\em $^{b}$ T-2, Theoretical Division, Los Alamos National Laboratory,
Los Alamos, NM 87545}\\
\end{center}
\begin{abstract}
Methods for radionuclide
production calculation in a high power proton
accelerator have been developed and applied to study production of
22 isotopes by high-energy protons and neutrons.
These methods are readily applicable to accelerator, and
reactor, environments other than the particular model we considered and to
the production of other radioactive and stable isotopes. We have also
developed methods for
evaluating cross sections from a wide variety of sources
into a single cross section set
and have produced an evaluated library covering about a third
of all natural elements. These methods also are applicable to an
expanded set of reactions.
A 684 page detailed report on this study,
with 37 tables and 264 color figures is available on the
Web at {\bf http://t2.lanl.gov/publications/publications.html},
or, if not accessible, in hard copy from the authors.
\end{abstract}

\begin{center}
{\large\bf 1. Introduction}\\
\end{center}

The widespread use of radionuclides in medical and industrial applications
is steadily increasing, leading suppliers to seek out new production
facilities. A reliable supply chain is necessary to both encourage new
applications and to replace aging production sources. The United States, 
in
particular, faces a domestic production shortfall.
It has been a policy of this country to import radioisotopes
from Canada and other countries.
The lack of a reliable supply has led to supply problems at times
that could be ameliorated by a domestic production facility. 

Among the possibilities
for radionuclide production are high power accelerators, either purpose
built or alongside existing applications. As an example of the latter, a
recent study by the Medical University of South Carolina
\cite{1} discussed the
production of medical radioisotopes at the proposed Accelerator Production
of Tritium Facility (APT)
\cite{apt}.
For instance, the comprehensive report \cite{1} states:
``The Committee on Biomedical Isotopes
(Institute of Medicine) concluded that a dedicated facility, such as the
proposed National Biomedical Facility (NBTF) which was a focus of their
report, is not economically justified as a source of radionuclides, whether
such a facility were a reactor or an accelerator \cite{a2}. However, 
nuclides
produced at a facility in operation primarily for other purposes would 
incur
only incremental costs and therefore be a more cost effective means of
production; contributions from the revenue stream of radionuclide sales
would partially, if not completely, offset the incurred incremental costs
of production."
The report \cite{1} considered a large number of
radioisotopes for present or future applications in medical treatment and
diagnostic procedures. Many of these radioisotopes could be produced in the
intense and energetic neutron and proton fluxes characteristic of the APT
target and blanket assembly.

We have undertaken a study to see to what extent existing nuclear data
models are applicable to calculations of radionuclide production in a high
energy, high power environment. We chose the APT target/blanket assembly as
a typical environment in which to study isotope production. In addition to
the availability of existing input models for our Monte Carlo flux
calculations, the high energies of the neutron and proton fluxes offer a
formidable test of the nuclear data. In a previous report
\cite{2}, we considered
the production of two radioisotopes -- $^{18}$F and $^{131}$I --
at two locations in the
APT blanket. We have extended that study to look at the production rates 
of
22 isotopes in nearly 500 locations throughout the APT target and blanket.
In addition to the 100 milliamp 1.7 GeV proton beam energy assumed in the
previous study, we also treat beam energies of 1.0, 1.2, 1.4, 1.6, and 1.8
GeV (all at 100 milliamps).

It should be noted that we have chosen the APT accelerator
for our
study as an example with which to
demonstrate the possibility of production of radioisotopes at such a
facility.
Radioisotopes can be produced also at other high power
accelerators, projected for the accelerator transmutation of nuclear
wastes (ATW) or other needs, such as the new spallation sources under
consideration in USA, Europe, Japan and elsewhere
\cite{gudowski}-\cite{hans98}, as well as at different
nuclear reactors. Our computational method is not limited to a
particular facility and can be used to study production
of radioactive as well as stable isotopes at any accelerator or reactor.
For practical reasons, the emphasis of our present study is on
radioisotopes.

The production rate of a radioisotope can be obtained from the integral of
the flux and cross section leading to the direct production of the
radioisotope as a reaction product. Additional production is realized from
other radionuclides that decay to the desired product. Evaluation of
production rates then requires knowledge of the neutron and proton fluxes 
at
some position in the production facility and cross sections leading to 
production
of the desired radionuclide and its progenitors. Ideally, one would use a
cross section library that includes data for all nuclides in the
neighborhood of the desired product in a transmutation code. Construction 
of
such a comprehensive library is beyond the scope of the present study.
Instead, we evaluated cross sections for reactions most likely to lead to
our desired products
\cite{lib98}.

We have prepared numerous figures and tables as part of this work. Space
available in this paper
prevents their inclusion
here. Our 684 page detailed report on this study \cite{medrep98},
with 37 tables and 264 color figures is available on the World Wide
Web
under: \\
{\bf http://t2.lanl.gov/publications/publications.html}.
We will provide also a link to exact location in:\\
{\bf http://www.rt66.com/~kvr/kvr\_pubs.html\#CONTRIBUTED},
and a limited number of
hard copies may be available from the authors.
\\

\begin{center}
{\large\bf  2. APT Modeling and Flux Tallies}\\
\end{center}

Figure 1 is a 3-dimensional rendering of a computer model of the
target/blanket assembly. The model is based on the Todosow geometry with a
16 cm $\times$ 160 cm beam area. The beam enters the assembly through the 
green
window on the right. The beam strikes tungsten-dominated ladders in the
center of the assembly; the position of these ladders correspond to the
pipes extending from the yellow manifolds. The lateral blankets extend to
either side of the beam line and ladder area. The downstream blanket
encompasses the region to the left of the ladder region, while the upstream
blanket is the narrow area between the entrance window and the ladder
region.

\begin{figure}[h!]
\hspace*{10mm}
\psfig{figure=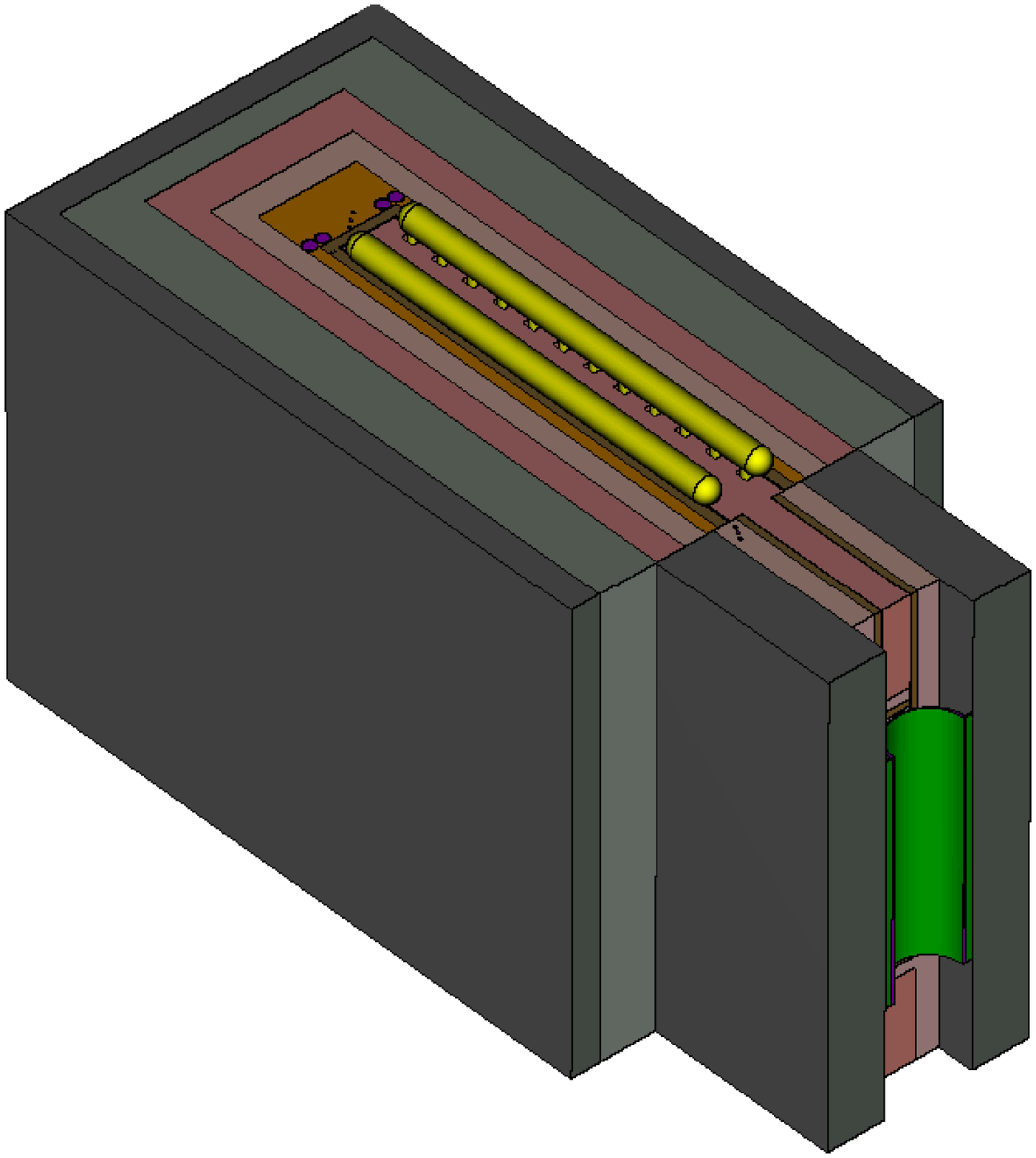,width=150mm,angle=0}
\end{figure}

{\small
Fig. 1. APT target/blanket assembly.\\
}

We used MCNPX version 2.1.1 to calculate neutron and proton fluxes
throughout the APT model, excluding pipes and ladders within the beam 
cavity
and some regions at the bottom of the model. We made runs for beam energies
of 1.0, 1.2, 1.4, 1.6, 1.7, and 1.8 GeV, assuming a 100 milliamp beam in
each case. For each case, the runs followed 120,000 incident protons.
\\

{\bf Tally Locations.}
Subdivision of cells in the upstream, downstream, and
lateral blanket regions and in the beam cavity between ladders yielded
approximately 183 cells in which the fluxes were tallied. Figure 2 shows 
the
cell locations in a slice in the X-Z plane through the middle of the target
model. We did not tally fluxes in the ladders (red areas in Figure 1), nor
in the coolant pipes (green areas).

Assuming reflective symmetry in the X direction (where Z is the beam
direction and Y is the vertical direction), tallies were taken over the
union of a cell on the -X side of the beam with its +X counterpart. Three
such pairs of symmetric cells are shaded light gray, yellow, and light blue
in Figure 2. Accounting for cells symmetric about X = 0, there were 102
tallies.
\\

\begin{figure}[h!]
\vspace*{-2.8cm}
\hspace*{10mm}
\psfig{figure=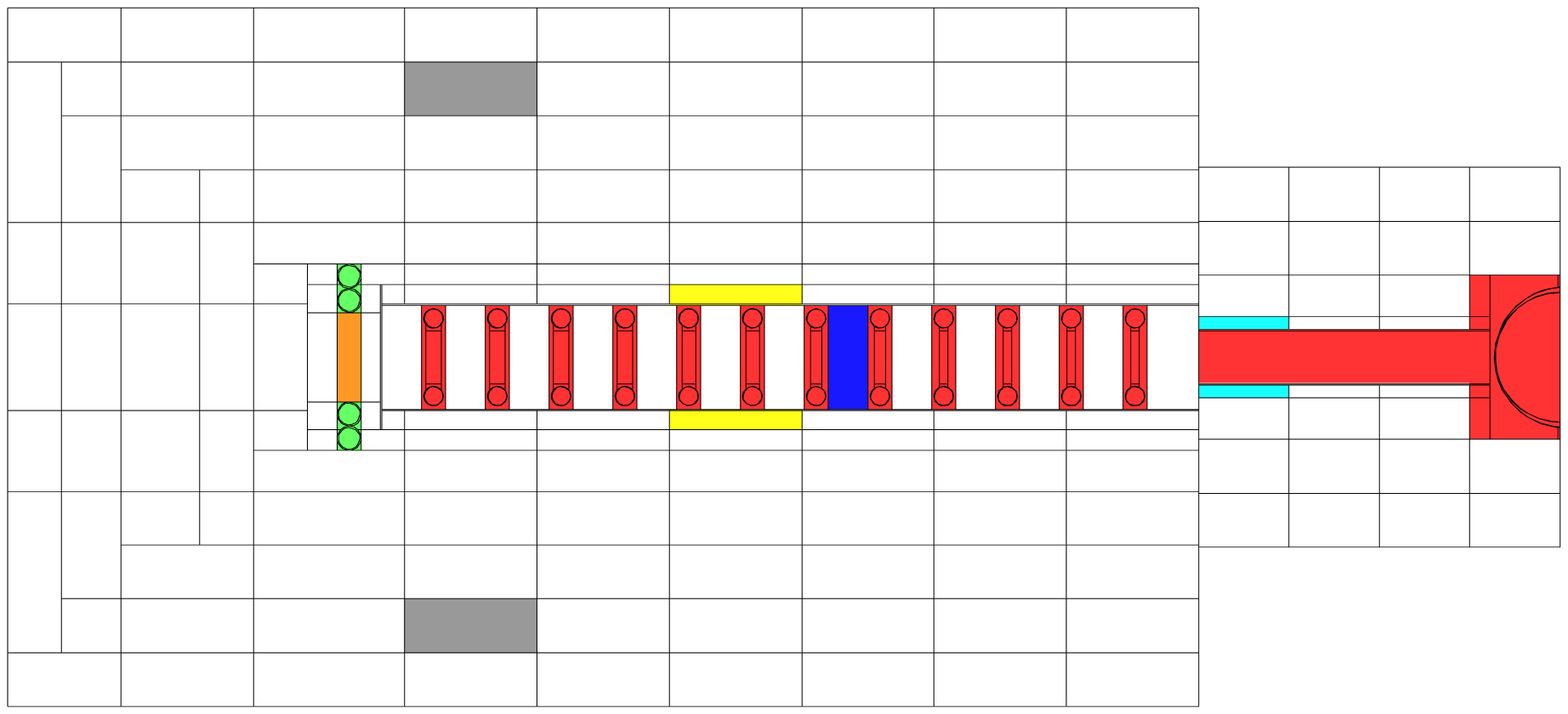,width=150mm,angle=0}
\end{figure}

\vspace*{-11.0cm}
{\small
Fig. 2. A slice through the middle of the APT target/blanket assembly showing
cells where tallies were taken.\\
}

{\bf Vertical Segments.}
We segmented the tallies into five equally spaced
vertical segments. Figure 3, a cross section through the upstream blanket,
shows the segmenting. Segment 1 is the topmost segment, Segment 3 is in the
middle position, and Segment 5 is the lowest. With a few exceptions, all
cells in the blankets extend the full height of the target, while the cells
in the beam line do not extend to the topmost and lowest segments. The
fluxes, and hence production rates, were greatest in the middle segment 
(3),
decreasing towards the top and bottom. Unless otherwise stated, the results
presented here are for the middle segment.
\\

{\bf Locations for Detailed Results.}
We present detailed summaries of the
production rates at four locations. Two of these locations, in the upstream
(light blue cells in Fig. 2) and downstream (orange cell) blankets, were
chosen to overlap the locations used in the previous study. The other two
locations, one behind the fifth ladder in the beam cavity (dark blue cell 
in
Fig. 2) and the other opposite the sixth, seventh, and eighth ladders in 
the
first lateral blanket row (yellow cells), are the locations of maximum
production rates for a great majority of reactions. For each reaction, we
found the cell with the maximum rate over our entire set of tallies, and 
the
cell with the maximum rate over all tallies excluding the beam cavity 
cells.
In approximately 80\% of the reactions, this procedure picked our two
selected locations. For all but a handful of the remaining reactions, the
picked cells were a neighbor of the selected locations.

We did not model any fixtures, such as irradiation tubes, that would be
required to produce radioisotopes in any location. Our results thus assume
any such fixtures would have no effect on the fluxes.

Spectrum plots of the neutron and proton fluxes in each segment of
the four selected locations are given in our detailed report
available on the Web \cite{medrep98}.
\\

\begin{figure}[t!]
\begin{minipage}[t!]{.30\linewidth} 
\psfig{figure=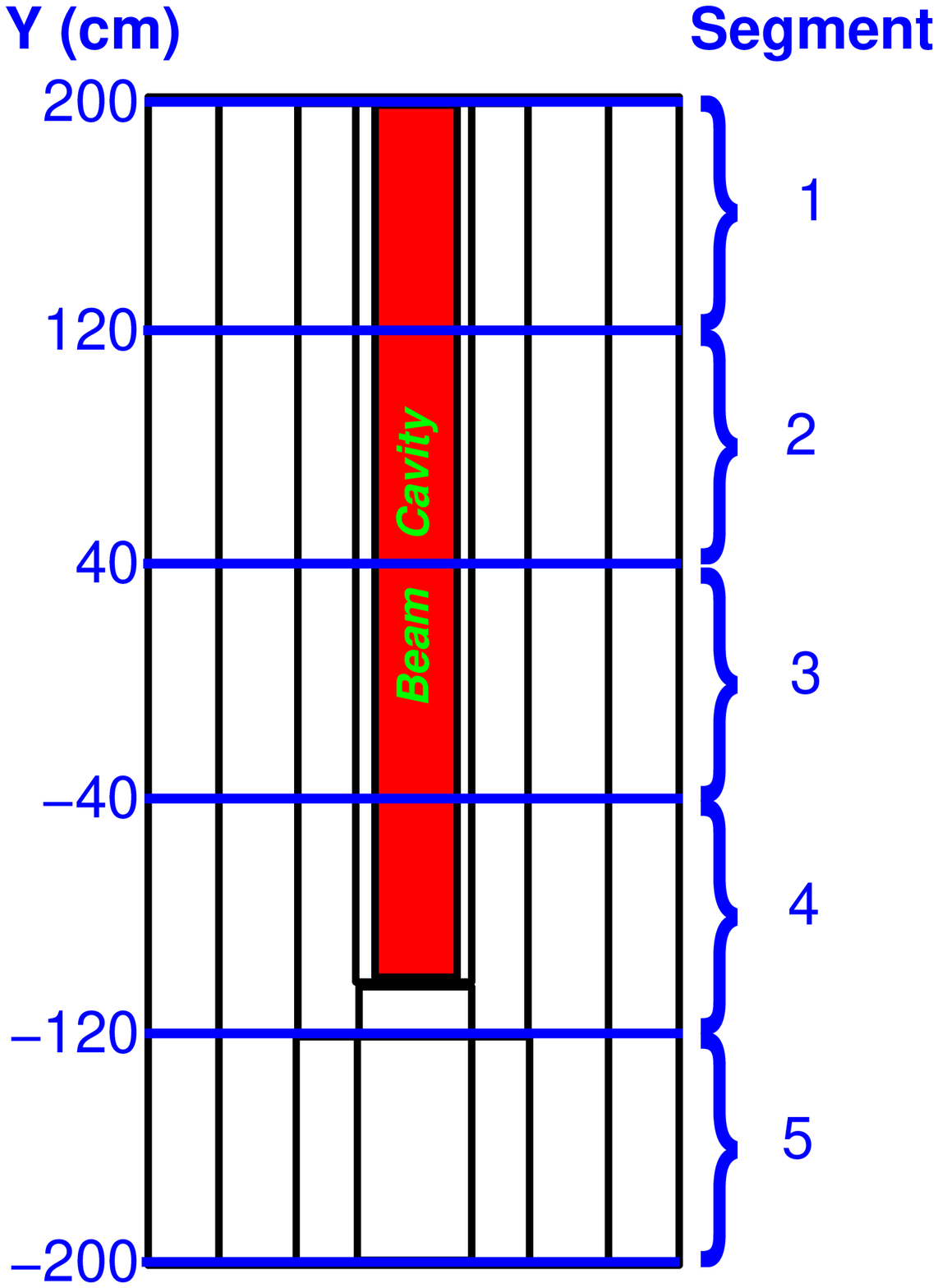,width=\linewidth,angle=0}
\end{minipage}\hfill
\begin{minipage}[t!]{.65\linewidth} 
{\small
Fig. 3. A cross section through the upstream section of the APT target/blanket 
assembly showing the vertical segment used in the tallies.
}
\end{minipage}
\end{figure}

{\bf Flux Color Contours on Planes.}
We prepared a variation of the target
geometry for display of the fluxes (and production rates) as color-coded
contours on a plane. We introduced 5 planes to represent the 5 vertical
segments; the vertical position of the plane lies at the center of the
corresponding segment. Each plane thus represents the flux averaged over a
segment. For visual orientation, we included the ladders, top manifolds, 
and
other elements in the revised geometry; these elements, shown in gray in
the plots, represent cells in which we did not take a flux tally. (This
revised model is used for visualization only; it was not used in the flux
calculations.)

Figure 4 shows a color contour plot of the neutron flux
for a 1.0 GeV beam energy.
The detailed Web report \cite{medrep98}
includes similar plots for neutron and proton flux contours at the 6 
energies
we consider.\\

{\large\bf 3. Cross Section Evaluations}\\

Our earlier work \cite{2}
involved the modeling of neutron- and proton-induced
reactions on isotopes of O, F, Ne, Na, Mg, Al, Xe, Cs, Ba and La for
energies to 1.7 GeV. We have now added reactions on isotopes of S, Cl, Ar,
K, Zn, Ga, Ge, As, Zr, Nb, Mo and Hg.

At present, neither available experimental data nor any
of the current models or phenomenological systematics
can be used alone to produce a reliable evaluated activation cross section
library covering a large range of target nuclides and incident
energies. Therefore, we chose to create our evaluated library \cite{lib98}
by constructing excitation
functions using all available experimental data along with calculations
using some of the more reliable codes, employing each of these sources in 
the
regions of
targets and incident energies where they are most applicable.
When we have reliable experimental data, they, rather than model results,
are taken as
the highest priority for our approximation.
Wherever possible, we attempted to construct a smooth transition from
one data source to another.

The recent {\em International Code Comparisons for Intermediate Energy 
Nuclear
Data} organized by NEA/OECD at Paris \cite{paris9497},
our own comprehensive benchmarks
\cite{2,lib98,medrep98,report97,act150},
and several studies by Titarenko et al. \cite{titarenko} have shown
that a modified version of the Cascade-Exciton model (CEM) \cite{cem}
as realized in the code CEM95 \cite{cem95}
and LAHET code system \cite{lahet,lahet283}
generally have the best predictive powers for
spallation reactions at energies above 100 MeV as compared to other
available models. 

\newpage
\begin{figure}[h!]

\hspace*{0mm}
\psfig{figure=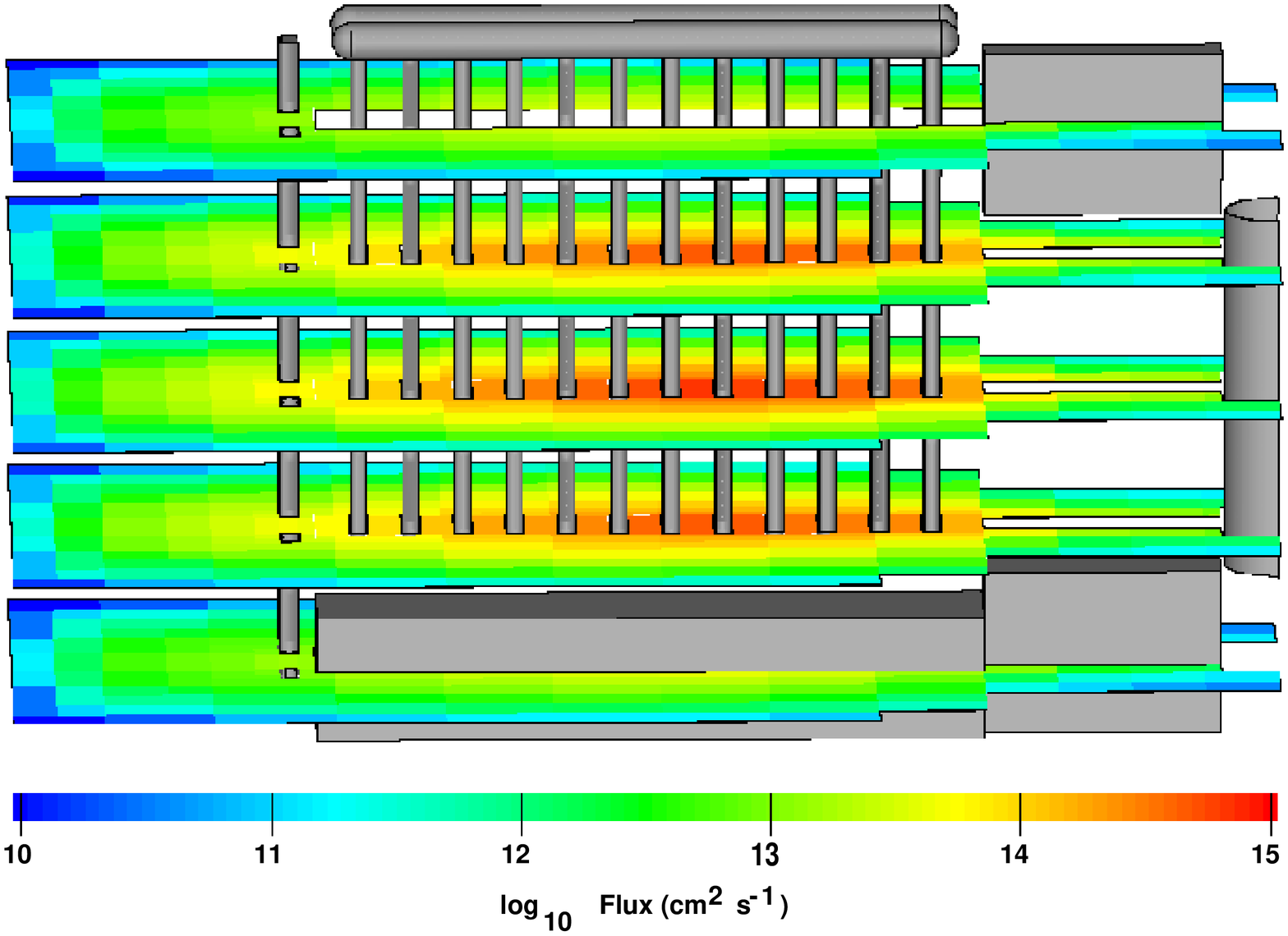,width=170mm,angle=-90}
\end{figure}
\vspace*{-1.0cm}

{\small
Fig. 4. Color contours of the neutron flux at a beam energy of 1.0 GeV.\\
}

Therefore, we choose
CEM95 \cite{cem95},
the recently improved version of the CEM code
\cite{cem98},
and LAHET (version 2.83) \cite{lahet,lahet283}
above 100 MeV to          
evaluate the required cross sections.
Specifically, we employ the calculated library described
in \cite{lib98}.
The same benchmarks have shown that at lower energies, the
HMS-ALICE code \cite{alice96}
most accurately reproduces experimental results as compared with other 
models.
We therefore use the activation library calculated by M. B. Chadwick 
\cite{mark}
with the HMS-ALICE code \cite{alice96} for protons
below 100 MeV and neutrons between 20 and 100 MeV. In the overlapping
region, between 100 and 150 MeV, we use both HMS-ALICE and CEM95 and/or
LAHET results. For neutrons below 20 MeV, we consider the data of the 
European
Activation File EAF-97, Rev. 1 \cite{9,eaf97} with some
recent improvements by
M. Herman \cite{herman}, to be the most accurate results available;
therefore we use
them as the first priority in our evaluation.

Measured cross-section data from our compilation described in \cite{lib98},
when available, are included together with theoretical results
and are used
to evaluate cross sections for study.
We note that when we put together all these different theoretical
results and experimental data, rarely do they agree perfectly with each
other, providing a smooth continuity of evaluated excitation functions.
Often, the resulting compilations show significant disagreement at energies
where the available data progresses from one source to another.
These sets are
thinned to eliminate discrepant data, providing data sets of more-or-less
reasonable continuity defining our evaluated cross sections \cite{lib98}
used here.

Examples with typical results of evaluated activation cross sections
for several proton reactions are shown in Fig. 5.
by broad gray lines. 51 similar color
figures for proton-induced reactions and 57 figures for neutrons,
can be found on the Web, in our detailed report \cite{medrep98}.

Our cross section tables reach to a maximum energy of 1.7 GeV. Between 1.7
and 1.8 GeV, we used the value at 1.7 GeV. Because most cross sections are
relatively independent of energy at these high energies, this assumption
should not be far removed from reality.\\

{\large\bf 4. Production Rate Calculations}\\

We consider production rates for the following 22 end product nuclides
of medical importance \cite{1}:

\begin{center}
\begin{tabular}{cccc}
$^{18}$F  & $^{35}$S & $^{89}$Sr & $^{133}$Xe \\

  $^{22}$Na   &    $^{67}$Cu   &    $^{89}$Zr   &    $^{131}$Cs \\

$^{32}$Si / $^{32}$P  &  $^{67}$Ga   &    $^{95}$Zr  &     $^{137}$Cs \\

   $^{32}$P    &    $^{68}$Ga  &     $^{95}$Nb   &   $^{193m}$Pt \\

   $^{33}$P  &  $^{68}$Ge / $^{68}$Ga  &  $^{131}$I   &   $^{195m}$Pt \\
\end{tabular}
\end{center}

To produce these nuclides, we calculated neutron and proton reactions on 
the
stable, naturally occurring isotopes of elements in the neighborhood of the
targets investigated. These 70 nuclides of 25 elements are:

\begin{center}
\begin{tabular}{cccccc}
 $^{18}$O  & $^{32}$S & $^{66}$Zn &  $^{89}$Y& $^{130}$Xe&$^{193}$Ir \\

      & $^{33}$S & $^{67}$Zn&   &    $^{131}$Xe  &                   \\

 $^{19}$F  & $^{34}$S & $^{68}$Zn & $^{90}$Zr& $^{132}$Xe&$^{197}$Au \\

     &  $^{36}$S&  $^{70}$Zn & $^{91}$Zr &$^{134}$Xe &               \\

$^{20}$Ne &   &    $^{92}$Zr &     & $^{136}$Xe&$^{196}$Hg           \\

$^{21}$Ne & $^{35}$Cl & $^{69}$Ga & $^{94}$Zr&      &$^{198}$Hg      \\

$^{22}$Ne & $^{37}$Cl & $^{71}$Ga & $^{96}$Zr& $^{133}$Cs&$^{199}$Hg \\

    &    &   &    &    &         $^{200}$Hg                          \\

$^{23}$Na&  $^{36}$Ar & $^{70}$Ge & $^{93}$Nb& $^{134}$Ba&$^{201}$Hg \\

     & $^{38}$Ar&  $^{72}$Ge &     & $^{135}$Ba&$^{202}$Hg           \\

$^{24}$Mg & $^{40}$Ar&  $^{73}$Ge & $^{92}$Mo& $^{136}$Ba&$^{204}$Hg \\

$^{25}$Mg&       & $^{74}$Ge & $^{94}$Mo &$^{137}$Ba &               \\

$^{26}$Mg  & $^{39}$K & $^{76}$Ge & $^{95}$Mo& $^{138}$Ba&           \\

      & $^{40}$K &     &  $^{96}$Mo &     &                          \\

      & $^{41}$K &      & $^{97}$Mo &$^{138}$La  &                   \\

$^{27}$Al &      & $^{75}$As & $^{98}$Mo &$^{139}$La &               \\

        &  &     &   $^{100}$Mo        &      &                      \\

\end{tabular}
\end{center}
\noindent{
For each reaction, we\\
}
\noindent{
1) constructed a continuous energy representation of the cross section
from the evaluation tables;}\\
2) formed a flux-weighted average cross section for each particle flux at
each location;\\
3) computed the one-hour irradiation end product $P$ production rate per
gram of target to each target nuclide / reaction product $p$ radionuclide
combination;\\
4) formed the one hour irradiation production rate per gram of target
nuclide or naturally occurring element.

\newpage
\begin{figure}[h!]
\hspace*{10mm}
\psfig{figure=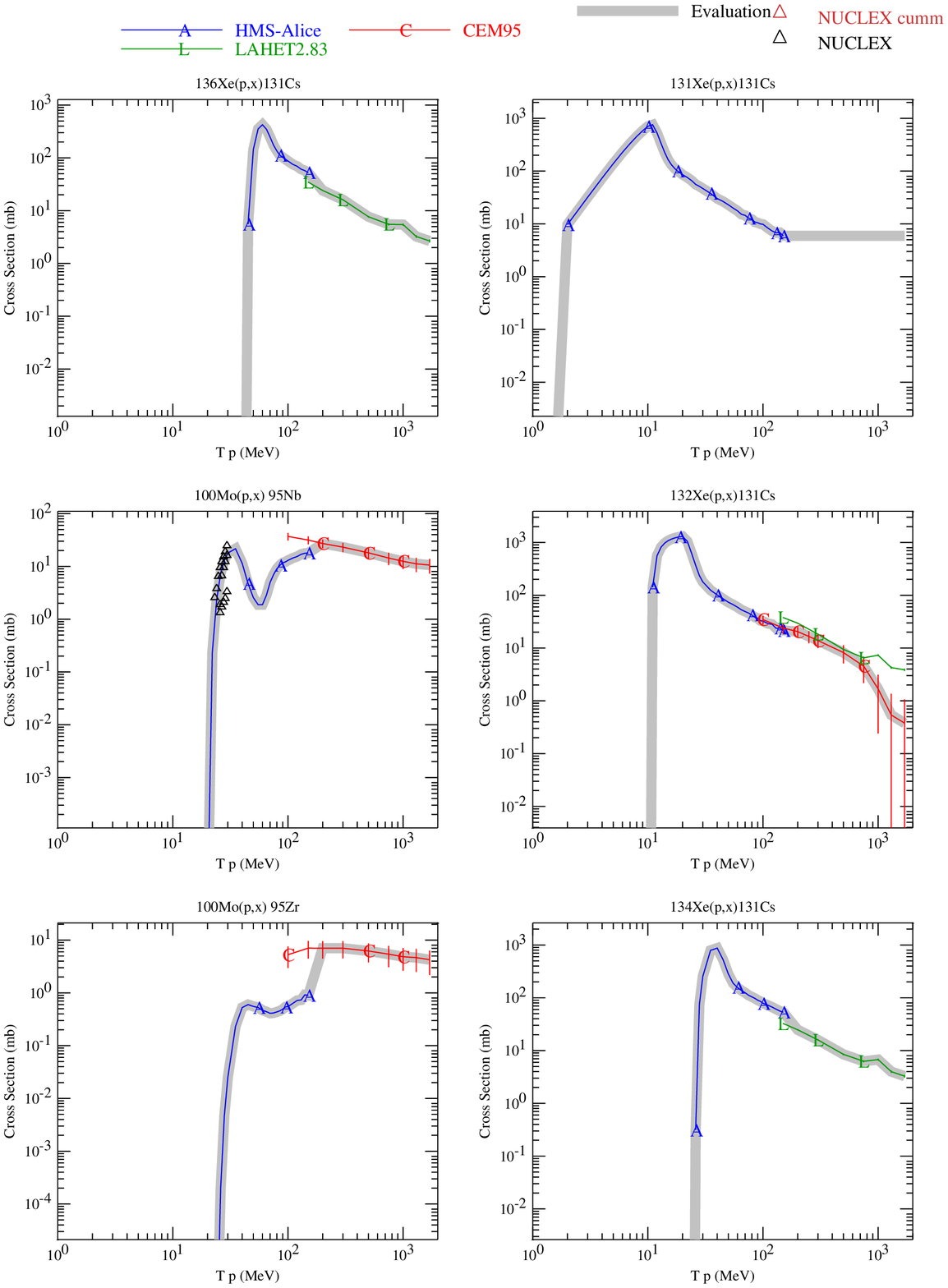,width=160mm,angle=0}
\end{figure}

{\small
Fig. 5. Examples of several evaluated proton-induced activation cross
sections. Evaluated cross sections are shown by broad gray lines,
other notations are given in the plots and described in the
text and in \cite{lib98}.  
}

\newpage

The flux-weighted cross section $\sigma_{tp}$
for each target $t$ and reaction product
$p$ is found by
$$\sigma_{tp} =
\frac{\int_{0}^{\infty} \phi (E) \sigma_{tp}(E) dE}
{\int_{0}^{\infty} \phi (E) dE}
= \frac{\sigma_{tp} \Phi} {\Phi} \mbox{ .}$$

\noindent{ 
The cross section $\sigma_{tP}$
 leading to end product $P$ is taken as the sum of all
cross sections for the direct production of $P$ and products $p$
decaying to $P$.}
The cross section $\sigma_{ZP}$ for element $Z$
leading to end product $P$ is obtained as
the natural-abundance-weighted sum of the cross sections
$\sigma_{tP}$ of the various
naturally occurring target nuclides of the element. For intermediate
nuclides that have multiple decay paths, we multiply the rate to account 
for
the branching factors to the desired end product.

The production rate $R_{tP}$ (Ci/g-hr) for each target $t$ --
end product $P$
combination is
$$R_{tP} = N_t \sigma_{tP} \Phi [ 1 - \exp (-\lambda_p T)]\mbox{ ,}$$

\noindent{
where $\lambda_p$ is the decay constant (s$^{-1}$) of end product
$P$, $T = 3600$s
corresponds to a one-hour irradiation,
$N_t = N_0/A_t$ is the atom density
(atoms/g) of the target material, $N_0$ is Avogadro's number
($6.022 \times 10^{23}$
atoms/mole), and $A_t$ is the atomic weight of the target.
$A_t$ is taken as the
integer mass number for isotopic targets and as the atomic weight for the
elements.\\

{\large\bf 5. Isotope Production Rates}\\

As an example, Table 1 shows a small part of our final results, for the
production rates of only $^{18}$F, for the upstream and downstream blanket
positions for a
1.7 GeV beam energy. This is just to have an idea how look our complete
tables in our Web report \cite{medrep98}
for the production all isotopes studied.
The Web version of the report includes similar but complete tables
for other beam energies. Also, on the Web are tables that gives detail
of these production rates, including the flux-averaged cross sections
and production rates for intermediate products,
and color contour plots
of production rates for natural element target at a beam energy of 1.7 GeV.
\\

\vspace*{-0.5cm}
\begin{center}
\begin{small}
Table 1\\ 
Production Rates in Forward and Downstream Blanket
Locations for a 1.7 GeV Beam Energy\\
\end{small}

\vspace*{0.2cm}

\begin{tabular}{cc|ccc|ccc}
\hline
 & &
\multicolumn{3}{|c} {Forward Blanket} &
\multicolumn{3}{|c} {Downstream Blanket} \\
Target & Product & Protons & Neutrons& Total& Protons& Neutrons& Total\\
Nuclide& Nuclide& Ci/g-hr & Ci/g-hr & Ci/g-hr & Ci/g-hr & Ci/g-hr & Ci/g-hr\\
\hline
$^{nat}$O &$^{18}$F &2.87E-06&0.00E+00&2.87E-06&1.66E-05&0.00E+00&1.66E-05\\
$^{nat}$F &$^{18}$F &7.96E-03&7.69E-02&8.48E-02&7.73E-02&2.23E-01&3.00E-01\\
$^{20}$Ne &$^{18}$F &5.16E-03&1.29E-02&1.81E-02&5.35E-02&7.85E-02&1.32E-01\\
$^{21}$Ne &$^{18}$F &2.91E-03&3.52E-03&6.44E-03&2.99E-02&2.86E-02&5.85E-02\\
$^{22}$Ne &$^{18}$F &2.47E-03&1.52E-03&3.99E-03&2.20E-02&1.45E-02&3.65E-02\\
$^{nat}$Ne&$^{18}$F &4.89E-03&1.17E-02&1.66E-02&5.03E-02&7.20E-02&1.22E-01\\
$^{nat}$Na&$^{18}$F &2.43E-03&1.02E-02&1.27E-02&2.41E-02&5.11E-02&7.52E-02\\
$^{24}$Mg &$^{18}$F &2.04E-03&4.29E-03&6.33E-03&2.21E-02&2.87E-02&5.08E-02\\
$^{25}$Mg &$^{18}$F &1.48E-03&1.95E-03&3.43E-03&1.68E-02&1.57E-02&3.24E-02\\
$^{26}$Mg &$^{18}$F &1.14E-03&9.73E-04&2.11E-03&1.26E-02&9.30E-03&2.19E-02\\
$^{nat}$Mg&$^{18}$F &1.88E-03&3.66E-03&5.54E-03&2.04E-02&2.51E-02&4.55E-02\\
$^{nat}$Al&$^{18}$F &1.15E-03&1.16E-03&2.31E-03&1.32E-02&9.67E-03&2.28E-02\\
\hline
\end{tabular}
\end{center}

\begin{center}
{\large\bf 6. Dependence on Position and Beam Energy}\\
\end{center}

We selected three groups of cells,
all in the middle segment, to explore the dependence of the production 
rates
on position in the target/blanket assembly and on the beam energy. Figure
6, a slice in the X-Z plane through the middle of the target, shows the
groupings. The {\em Beam Cavity Cells}
are light blue, the {\em Downstream Cells} are
dark blue, and the{\em  Lateral Cells}
are yellow. The orange cell belongs to both
the cavity and lateral groups.
\\

\begin{figure}[h!]
\hspace*{0mm}
\psfig{figure=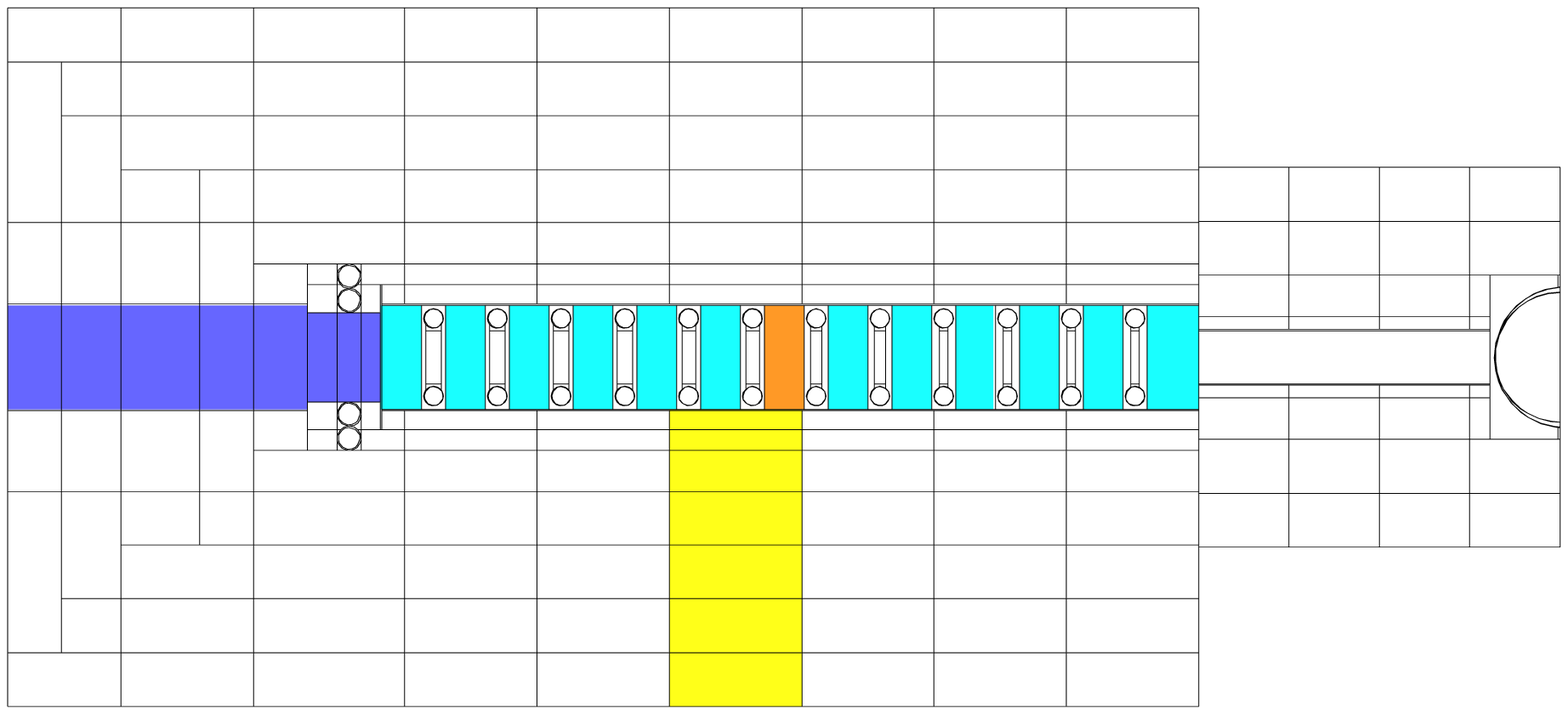,width=115mm,angle=0}
\end{figure}
\vspace*{-4.3cm}

{\small
Fig. 6. A slice through the middle of the APT target/blanket assembly
showing the locations of the cell grouping in the energy and distance 
dependence studies.\\
}

For the beam cavity cells, the distance is measured along the beam starting
at the first upstream light blue cell. For the downstream cells, the
distance is measured from the end of the beam cavity. The lateral distance
is from the centerline of the beam cavity towards the outside of the
assembly.

Figures showing the variation of production rates with beam energy and
distance from natural
element targets are included in the Web version of the report.
The rates increase with beam energy, by between factors of 1.5 and 5 from
beam energies of 1.0 to 1.8 GeV. The downstream cells closest to the beam
cavity are less sensitive to beam energy than those further downstream.
There is less spread in the beam energy dependence in the lateral cells  
than
in the other groups. Within the beam cavity, the rates peak at a distance  
of
140 cm. The rates decrease exponentially with distance into the lateral
blanket. Downstream of the first few cells in the downstream blanket, the
rates also decrease exponentially with distance into the downstream
blanket.\\

\newpage
\begin{center}
{\large\bf 7. Summary}\\
\end{center}

We have characterized the radiation environment in a high power proton
accelerator and developed methods for radionuclide production calculations.
These methods are readily applicable to accelerator,  and
reactor,  environments other than the particular model we considered and to
the production of other radioactive and stable isotopes.
We have also
developed methods for
evaluation cross sections from a wide variety of sources
into a single cross section set. These methods also are applicable to an
expanded set of reactions.

While the agreement among the different cross section models and with
experimental data is good for quite a few reactions, a significant number 
of
reactions remain problematical. A particular problem is the lack of high
energy data for $^{193m}$Pt and $^{195m}$Pt production.
We will look into the
possibility of extracting excited state production from LAHET and other
calculations in future work.

Commenting on the feasibility and economic viability of the production of
any particular isotope is beyond the scope of this work. Such information,
which must come from experts in the radionuclide arena, will be a vital
ingredient in choosing which cross sections should be subjected to greater
scrutiny.

This study of isotope production in targets within an
accelerator spallation target/blanket assembly is
particular to the APT --- now the backup concept for 
tritium production in the US DOE. Because of the broad
range of beam energy and target positioning in the study,
as well as the spatial mapping of neutron and proton fluxes and 
isotope production throughout the geometry, results are
applicable to a wide range of beam energies and beam intensities
of spallation targets now under consideration in the USA, Europe, 
Japan and elsewhere.
\\

\begin{center}
{\bf Acknowledgements}\\
\end{center}

We thank many colleagues, in particular M. B. Chadwick and A. J. Sierk,
for fruitful collaboration with us at different stages of this work.
We express our gratitude to  R.~E.~MacFarlane and
L.~S.~Waters for interest in and support of the present work.
This study was partially supported by the U.~S.~Department of Energy.

\end{document}